\documentclass[aps,prl,superscriptaddress,twocolumn,showpacs]{revtex4}
\usepackage{amsmath}
\usepackage{graphicx}
\usepackage{amssymb}
\usepackage{bbm, color, soul}
\usepackage{graphicx}
\usepackage{adjustbox}
\usepackage{mathtools}
\usepackage{booktabs}

\usepackage{color}

\usepackage{tikz}
\usetikzlibrary{positioning}

\usepackage{amsthm}
\usepackage{tikz-cd}
\usepackage[all]{xy}
\usepackage{amsfonts}
\usepackage{mathrsfs}
\usepackage{hyperref}
\usepackage{MnSymbol} 

\usepackage{amsmath,amssymb,amsfonts}
\newcommand{\C}{\mathbb{C}}

\theoremstyle{plain}

\theoremstyle{definition}
\newtheorem{defn/}[equation]{Definition}

\theoremstyle{remark}
\newtheorem{rmk/}[equation]{Remark}
\newtheorem*{rmk*}{Remark}

\newtheorem*{claim*}{Claim}

\newcommand{\CC}{\mathbb{C}}
\newcommand{\DD}{\mathbb{D}}

\newcommand{\ZZ}{\mathbb{Z}}

\begin{document}

\title[]{Singularity Selector: Topological Chirality via Non-Abelian Loops around Exceptional Points}
\author{Kyu-Won \surname{Park}}
\affiliation{Department of Mathematics and Research Institute for Basic Sciences, Kyung Hee University, Seoul, 02447, Korea}

\author{KyeongRo \surname{Kim}}
\email{kyeongrokim14@gmail.com}
\affiliation{Research Institute of Mathematics, Seoul National University, Seoul 08826, Korea}

\author{Kabgyun \surname{Jeong}}
\email{kgjeong6@snu.ac.kr}
\affiliation{Research Institute of Mathematics, Seoul National University, Seoul 08826, Korea}
\affiliation{School of Computational Sciences, Korea Institute for Advanced Study, Seoul 02455, Korea}

\date{\today}

\begin{abstract}
Chirality is more than a geometric curiosity; it governs measurable asymmetries across nature, from enantiomer-selective drugs and left-handed fermions in particle physics to handed charge transport in Weyl semimetals.
We extend this universal concept to non-Hermitian systems by defining topological chirality, an invariant that emerges whenever an exceptional-points (EP) pair is present. Built from the non-commutative fundamental group and its braid representation, topological chirality acts as a singularity selector: clockwise EP loops occupy a homotopy class that avoids EPs, whereas counter-clockwise mirrors are equivalent only if they cross the EPs themselves. We confirm this binary rule in an optical microcavity and a non-Hermitian topological band. The same two-sheeted topology governs EP pairs in spin systems, photonic crystals and hybrid light–matter structures, where EP encirclements have already been demonstrated, so the framework transfers without alteration and confirms its experimental viability. Our findings lay the cornerstone for interpreting loop-sensitive observables such as spectral vorticity, the complex Berry phase and the non-Abelian holonomy. Finally, a gluing-of-planes construction extends the invariant to an $n$-sheeted surface hosting 2m EPs, unifying higher-order EP pairs.
\end{abstract}

\maketitle

\section{Introduction}
Chirality that is the inability of an object to coincide with its mirror image is not just a geometric curiosity; it is a cornerstone from  molecular biology to modern physics.  It dictates drug efficacy, governing about half of all pharmaceuticals~\cite{Shen2013}, and drives the chiral anomaly in Dirac and Weyl semimetals~\cite{Xiong2015Science,Xu2015Science}.  Structured chiral light enables enantio-selective sensing and robust optical control~\cite{Smirnova2024NatPhoton,Mun2020LSA}, while molecular handedness yields spin filtering~\cite{Pacchioni2023NatRevMat}.  Hyper-Raman techniques have recently pushed optical activity into new territory~\cite{Jones2024NatPhoton}.  These advances underscore that harnessing chirality is essential not only for applications but also for probing the deepest symmetries and asymmetries of nature.

In non-Hermitian (NH) settings, chirality takes on new forms driven by exceptional points (\textsf{EP})s where eigenvalues and eigenvectors coalesce~\cite{R09, W04, T66} leading to phenomena such as spatial mode imbalance~\cite{Wiersig2014a, Schomerus2014}, local spatial chirality~\cite{Yi2018PRL, Kullig2019}, and encircling-induced asymmetric switching~\cite{DM16, XA22, Zhang2019}. These previous chirality depend sensitively on the details of the eigenmodes or the precise encirclement path.

Here, we introduce a fundamentally different form of NH chirality-\emph{topological chirality} that originates from the {\em non-commutativity} of \textsf{EP} loops themselve~\cite{Zhong2018NatCommun, Guria2024NatCommun}.  The key is that two loops which encircle the same pair of \textsf{EP}s the same number of times (for instance, the sequence $ab$ versus $ba$) belong to distinct homotopy classes and cannot be bi-continuously deformed into one another. Defined via the non-commutative orbifold fundamental group~\cite{Thurston1997} and its surface braid group representation~\cite{B74}, topological chirality partitions every \textsf{EP} encirclement into two inequivalent classes: loops and their mirror images, thereby endowing chiral responses with true topological protection, independent of any specific eigenmode or loop geometry.

To demonstrate and exploit this notion, we perform numerical simulations in two physically distinct platforms: an optical microcavity~\cite{Park2018PRE, Park2020SciRep} and a non-Hermitian Dirac Hamiltonian~\cite{BergholtzReview, Zhen2015Nature, Zhou2018}. The resulting binary classification can be directly imprinted on loop sensitive observables such as spectral winding number~\cite{BergholtzReview,Shen2018PRL, Zhang2020PRL}, the complex Berry phase~\cite{
UzanNarovlansky2024Nature, Singhal2023PRResearch}, and the non-Abelian holonomy matrix~\cite{Chen2025NatCommun, Shan2024PRL}.  Moreover, this \textsf{EP} pair is abundance including  spin systems~\cite{Galda2019SciRep}, photonic crystals~\cite{Nguyen2023APL}, and hybrid light-matter structures~\cite{Su2021SciAdv}. Thus, our work  can be applicable equally to such systems.  Finally, we present a gluing of planes construction that can extends topological chirality to an $n$-sheeted Riemann surface hosting $2m$ \textsf{EP}s, thereby establishing a unified algebraic topological framework for higher order NH degeneracies and sharpening the theoretical foundations of NH topology.

\section{Chirality in Non-Hermitian Systems}

Building on previous studies, we classify chirality in non-Hermitian systems into three principal classes, each reflecting a distinct physical origin:

\paragraph{(i) Spatial mode imbalance.}
In many optical resonators such as slightly deformed microdisk cavities or coupled resonator waveguides, the ideal clockwise (CW) and counter-clockwise (CCW) whispering-gallery modes become weakly coupled by surface roughness or refractive-index perturbations, leading to backscattering between the two propagation directions.  A convenient way to quantify the resulting directional bias is
\begin{align}
  \alpha = \frac{|A| - |B|}{|A| + |B|} \;\in\; [-1,1],
\end{align}
where \(A\) and \(B\) are the CW\(\to\)CCW and CCW\(\to\)CW scattering amplitudes.  Physically, \(\alpha=0\) indicates that scattering is reciprocal (no net circulation), while \(\alpha\to\pm1\) corresponds to one-way (chiral) propagation, as exploited in unidirectional microlasers and nonreciprocal photonic devices.\cite{Wiersig2014a,Schomerus2014}

\paragraph{(ii) Local spatial chirality.}
Near an \textsf{EP}, the two coalescing eigenmodes lock with a relative phase \(\pm\pi/2\), producing an intrinsic “polarization chirality,” independent of any global CW/CCW amplitude ratio.  This phase locking can manifest as a handed vortex in the local current (or Poynting vector),
\begin{align}
  \mathbf{J}(\mathbf{r})
  = \Im\bigl[\Phi^*(\mathbf{r})\,\nabla\Phi(\mathbf{r})\bigr],
\end{align}
where:
 \(\Phi(\mathbf r)=\psi_1(\mathbf r)\pm i\,\psi_2(\mathbf r)\) is the complex eigenmode (with real components \(\psi_1,\psi_2\)),
 \(\Im[\cdot]\) denotes the imaginary part,
 \(\nabla\) is the spatial gradient operator.

Here \(\mathbf J\) circulates around the point where the field amplitude \(\lvert \Phi(\mathbf r)\rvert\) vanishes, forming a vortex irrespective of the CW/CCW amplitude ratio\cite{Yi2018PRL,Kullig2019}.

\paragraph{(iii) Dynamical encircling–induced asymmetric switching.}
 Encircling of an \textsf{EP} in non-Hermitian systems inevitably violates the conventional adiabatic theorem: the closing of the spectral gap at the EP forces non-adiabatic transitions (NATs), which are both \emph{discontinuous} and \emph{path-dependent}.  These NATs give rise to characteristic asymmetric mode switching: clockwise versus counter-clockwise loops yield distinct final eigenmodes.  Because the transitions depend sensitively on the exact speed and trajectory, no robust, loop-independent (i.e., topologically protected) behavior emerges\cite{DM16,XA22,Zhang2019}.

\medskip
Together, these three mechanisms—mode imbalance, local spatial chirality, and dynamical encircling–induced asymmetric switching—constitute the conventional toolkit for identifying chirality in non-Hermitian systems.  However, this toolkit depends on specific mode structures or loop geometries, resulting in a lack of any topological protection.  In this paper, we introduce a fundamentally distinct, \emph{topological chirality}, driven by the non-commutative orbifold fundamental group and its surface braid-group representation, which is \emph{not only} path-independent but \emph{also} entirely independent of the underlying mode structure—thereby endowing chirality with true, robust topological invariance.```

\begin{figure*}[ht]
  \centering
  \includegraphics[width=\textwidth]{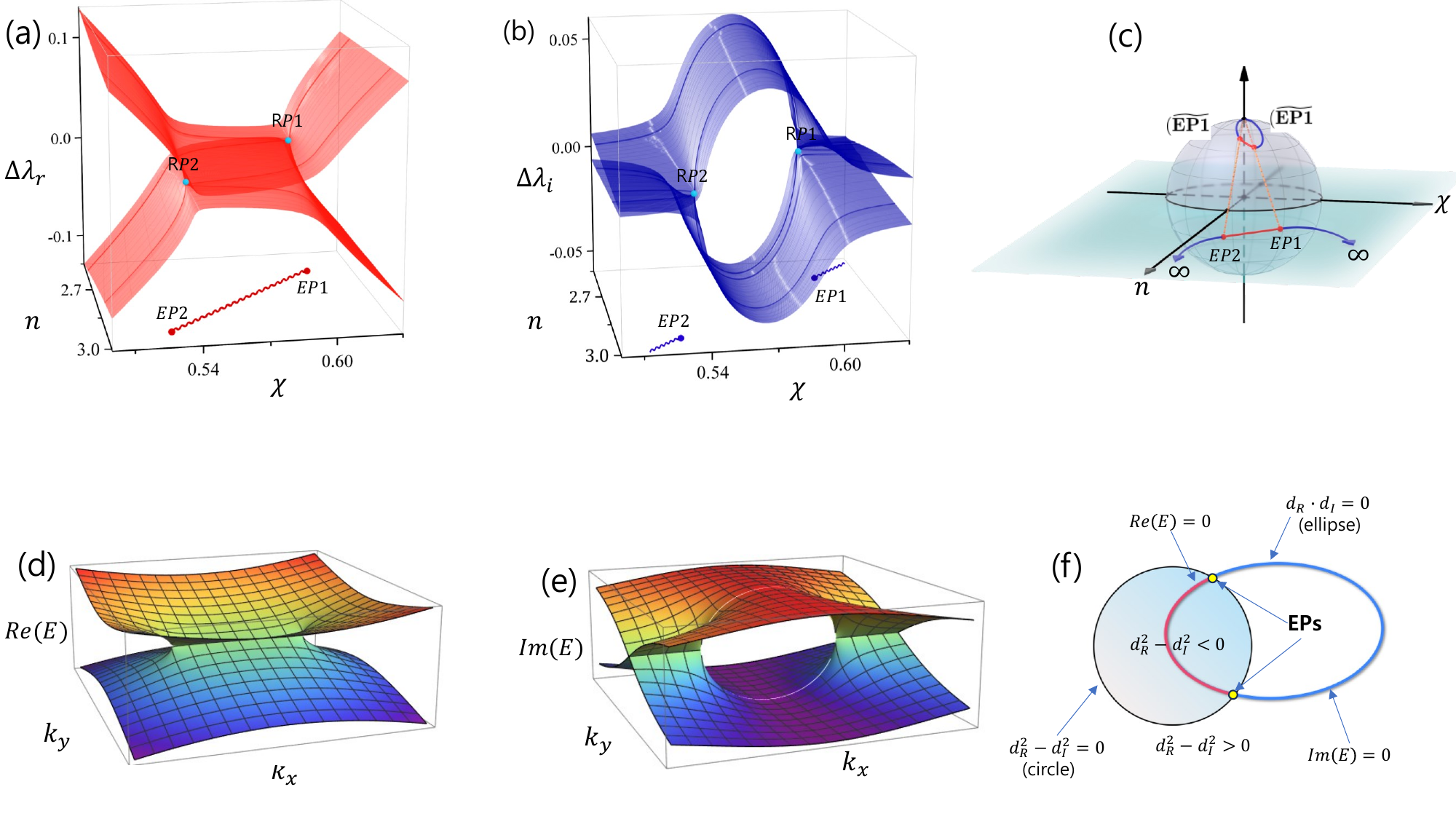}
  \caption{%
    \textbf{Universal appearance of an \textsf{EP} pair in two disparate open systems.}
    (a, b) Real and imaginary parts of the eigenvalue surface for an \emph{elliptical optical microcavity} as the deformation parameter \(\chi\) and interior refractive index \(n_{\mathrm{in}}\) are varied.  The two coloured sheets touch at precisely two isolated degeneracies (blue dots), forming an \textsf{EP} pair.
    (c) Same data mapped to the Bloch (Riemann) sphere: See the details in supplemantary.
    (d, e, f) Real and imaginary parts of the spectrum for a \emph{non-Hermitian Dirac model}
    \(H = k_x\sigma_x + k_y\sigma_y + i\sigma_x\); the control parameters are the crystal momenta \((k_x,k_y)\).  Again two \textsf{EP}s appear in the parameter space.
    (f) Parameter space view of the lattice model: the loci defined by
    \(d_R^{2}-d_I^{2}=0\) (circular curve) and
     \(d_R\!\cdot\!d_I = 0\)(elliptic curve); their intersections are the \textsf{EP} pair.  In regions where one condition holds and the other does not, the spectrum exhibits “Fermi‐arc” behavior: when
    \(d_R\!\cdot\!d_I = 0\) and \(d_R^2 - d_I^2 < 0\) (shaded area), the real part of the energy vanishes (pink curve); conversely, when
    \(d_R\!\cdot\!d_I = 0\) and \(d_R^2 - d_I^2 > 0\) (outside shaded area), the imaginary part vanishes (sky blue curve).
    Panels (a)-(c) and (d)-(f) thus visualise the \textsf{EP} pairs described by $f(z)=\sqrt{(z-z_{1})(z-z_{2})}$ realised in an optical microcavity and \emph{non-Hermitian Dirac model}, underscoring the abundance and universality of \textsf{EP} pairs.
  }
  \label{Figure-1}
\end{figure*}

\section{Abundance of \textsf{EP} Pairs}
Non-Hermitian (NH) behavior arises whenever a physical system is open to its environment, allowing energy or particles to leak, gain, or scatter into uncontrolled channels.  In many experiments two modes dominate the dynamics, so it suffices to describe the system with an effective two-level Hamiltonian:
\begin{align}
H_{\mathrm{NH}}
   =
   \begin{pmatrix}
     \lambda^{1} & g \\[4pt]
     g           & \lambda^{2}
   \end{pmatrix},
\end{align}
where the complex parameters \(\lambda^{1,2}\) include self-energy shifts and radiative losses, and \(g\in\mathbb{C}\) encodes the coherent + dissipative coupling between the two states.\cite{R09,W04,T66}  The eigenvalues are
$
\lambda_{\pm}
   =
   \lambda_{\mathrm{AV}}
   \pm
   \sqrt{\Delta^{2}+g^{2}},
$
with
$
\lambda_{\mathrm{AV}} = \tfrac{1}{2}(\lambda^{1}+\lambda^{2}),
\qquad
\Delta = \tfrac{1}{2}(\lambda^{1}-\lambda^{2}).
$
\textsf{EP}s occur when \(\Delta^{2}+g^{2}=0\), causing both eigenvalues and eigenvectors to coalesce.

Specifically, in the language of topological band theory, degeneracies such as Weyl or Dirac points are protected by symmetry and dimensional constraints. In a Hermitian two‐band model, real eigenvalue crossings occur only at isolated points where two independent real equations are satisfied. Non-Hermitian \textsf{EP}s generalize this notion by coupling both the real and imaginary parts of the Hamiltonian.

The simplest and most illuminating example is the generic two band non-Hermitian Pauli Hamiltonian in momentum space~\cite{BergholtzReview}:
\begin{align}
H(\mathbf{k})
  = \mathbf{d}_R(\mathbf{k})\!\cdot\!\boldsymbol\sigma
  + i\,\mathbf{d}_I(\mathbf{k})\!\cdot\!\boldsymbol\sigma
  + d_{0}(\mathbf{k})\,\sigma_{0}
  \label{eq:EPconditions}
\end{align}

where \(\mathbf{d}_R,\mathbf{d}_I\in\mathbb{R}^3\), \(\boldsymbol\sigma=(\sigma_x,\sigma_y,\sigma_z)\) are the Pauli matrices, and any \(d_0\,\sigma_0\) term (with \(\sigma_0\) the \(2\times2\) identity) has been set to zero.  Its complex eigenvalues read
\begin{align}
E_{\pm}(\mathbf{k})
  = \pm\sqrt{\;\lvert\mathbf{d}_R\rvert^{2}
               - \lvert\mathbf{d}_I\rvert^{2}
               + 2\,i\,(\mathbf{d}_R\!\cdot\!\mathbf{d}_I)\;}\,,
\end{align}
and \textsf{EP}s occur when the radicand vanishes, i.e.\ \(\lvert\mathbf{d}_R\rvert^{2}-\lvert\mathbf{d}_I\rvert^{2}=0\) and
\(\mathbf{d}_R\!\cdot\!\mathbf{d}_I=0\). Thus, unlike Hermitian degeneracies (which require tuning all three real components of \(\mathbf{d}_R\)), non-Hermitian \textsf{EP}s of codimension 2 generically appear as stable nodal points in two-dimensional parameter spaces and only form nodal lines when embedded in three or more dimensions without additional fine-tuning.

\subsection{\textsf{EP} pairs in an open optical microcavity}
Figures~\ref{Figure-1}(a)-(c) illustrate a two-dimensional elliptical microcavity with semi-axes
\(\,a=1+\chi\) and \(b=1/(1+\chi)\), keeping the area fixed at \(\pi\).
The exterior refractive index is \(n_{\mathrm out}=1\), while \(n_{\mathrm in}\) and the deformation \(\chi\) are varied.
Transverse-magnetic modes satisfy
\begin{align}
  \nabla^{2}\psi + n^{2}k^{2}\psi = 0,
\end{align}
which we solve using the boundary-element method\cite{Wiersig2003}.
Outgoing-wave boundary conditions render \(k = k_r + i k_i\) complex. Panels~(a) and (b) plot the relative splittings $
  \Delta\lambda_r = \Re\bigl(\lambda_{\pm}-\lambda_{\mathrm{AV}}\bigr),
  \quad
  \Delta\lambda_i = \Im\bigl(\lambda_{\pm}-\lambda_{\mathrm{AV}}\bigr),$
as functions of \(\chi\) and \(n_{\mathrm in}\), where \(\Delta\lambda_{\pm}=\lambda_{\pm}-\lambda_{\mathrm{AV}}\).

The corresponding Riemann surface has ramification points (blue dots), which map to \textsf{EP}s in the \((\chi,n_{\mathrm in})\) parameter plane.  These \textsf{EP}s act as branch points connected pairwise by branch cuts.  To visualize this topology, we project the extended parameter space onto the Riemann sphere in Fig.~\ref{Figure-1}(c) (see Supplementary Section 1).


\subsection{\textsf{EP} pairs in a non-Hermitian two-band model}
\label{subsec:NHDirac}
To emphasise the universality of Eq.~\ref{eq:EPconditions}, we examine the
\emph{non-Hermitian Dirac Hamiltonian}
\begin{equation}
H(\mathbf k)=k_x\sigma_x+k_y\sigma_y+i\,b_x\sigma_x,
\qquad b_x\in\mathbb R ,
\end{equation}
defined on the two-dimensional Brillouin zone (BZ) spanned by
$(k_x,k_y)$. The Hermitian part $k_x\sigma_x+k_y\sigma_y$ reproduces the familiar Dirac cone of graphene and other two-dimensional Dirac lattice models,
whereas the anti-Hermitian term $i\,b_x\sigma_x$ adds a uniform gain-loss imbalance along the $\sigma_x$ pseudospin axis.

Writing $H=\mathbf d_R\!\cdot\!\boldsymbol\sigma
        +i\,\mathbf d_I\!\cdot\!\boldsymbol\sigma$
with $\mathbf d_R=(k_x,k_y,0)$ and $\mathbf d_I=(b_x,0,0)$, the \textsf{EP} conditions $d_R^{2}=d_I^{2}$ and $\mathbf d_R\!\cdot\!\mathbf d_I=0$
(cf.\ Eq.~\ref{eq:EPconditions}) reduce to $ k_x = 0,k_y = \pm b_x .$
Simultaneously satisfying both constraints therefore selects two isolated momenta $\mathbf k=(0,\pm b_x)$ inside the BZ: an \textsf{EP} pair that
serves as the non-Hermitian counterpart of the single Dirac point in a Hermitian system. Choosing the representative value $b_x = 1$ places the pair at
$\mathbf k = (0,\pm1)$. Figures~\ref{Figure-1}(d)–(f) show the corresponding real and imaginary energy sheets together with the intersecting constraint curves.
The surfaces were obtained by direct diagonalisation of $H(\mathbf k)$ in \texttt{Wolfram~Mathematica~14};
These visuals reproduce up to a uniform rescaling the archetypal \textsf{EP} pair depicted in Fig.~3 of Ref.~\cite{BergholtzReview}.


\subsection{Unified perspective}
Although the elliptical microcavity and the non-Hermitian Dirac Hamiltonian operate in completely different physical regimes-optical resonance modes versus lattice Bloch electrons-their spectra share the same analytic shape:
\begin{align}
f(z)=\sqrt{(z-z_{1})(z-z_{2})}.
\end{align}
The two branch points \(z_{1,2}\) map directly onto the numerically identified \textsf{EP}s in Fig.\,\ref{Figure-1}(a-c) for the microcavity and Fig.\,\ref{Figure-1}(d-f) for the Dirac model.

\textsf{EP} pairs described by this square-root form are not confined to these two platforms.  They also appear in spin systems~\cite{Galda2019SciRep}, photonic crystals~\cite{Nguyen2023APL}, and hybrid light-matter structures~\cite{Su2021SciAdv}. Because the underlying two-sheeted Riemann topology for the \textsf{EP} pair is identical, the framework developed here can be transferred to any of these settings without modification.

\section{Fundemantal group associated with encircling of \textsf{EP} pairs}
\begin{figure*}
\centering
\includegraphics[width=\textwidth] {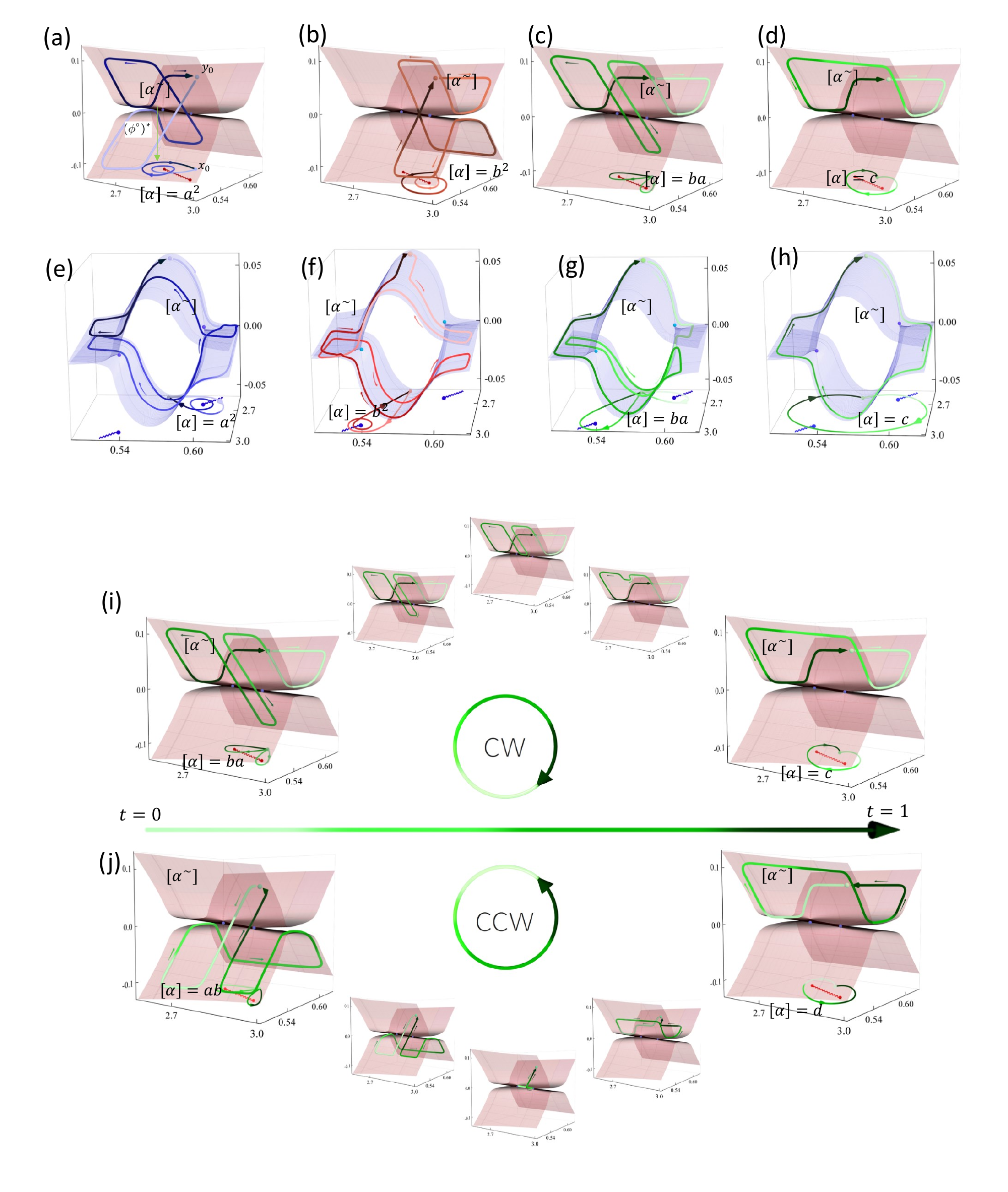}
\caption{
\textbf{Upper panels (a)-(h)}:
Panels (a)-(d) depict representative loops $[\tilde\alpha]\in\pi_{1}(Y,y_{0})$ on the Riemann surface encircling $RP_1$, $RP_2$, both points individually, and both simultaneously.  Panels (e)-(h) show their projected loops $[\alpha]=(\phi^{\circ})_{*}[\tilde\alpha]\in\pi_{1}(X,x_{0})$ around the corresponding exceptional points $\textsf{EP}_1$, $(\textsf{EP})_2$, etc.
\textbf{Lower panels (i),(j)}:
Homotopy-lifting diagrams for CW encirclement (i) and CCW encirclement (j), illustrating the bijection between loops on $X$ and their lifts on $Y$ as time t increase.}

\label{Figure-2}
\end{figure*}


Encircling exceptional points is central to non-Hermitian physics: it permutes eigenstates and imprints a complex Berry phase.
Moreover, such looping protocols reveal the underlying topology of the associated Riemann surfaces, providing insight into branch point connectivity and sheet structure. The present section formalises this insight by mapping EP encirclement to a problem in algebraic topology, specifically, to the fundamental group of a twice-punctured plane and its lift to a two-sheeted Riemann surface, thereby providing a rigorous framework for all subsequent calculations and simulations.

Viewing the real two-dimensional control-parameter plane $\mathbb R^{2}$ as its complexification $\mathbb C$, removing the two exceptional points yields the twice-punctured complex plane
$X=\mathbb C\setminus\{\textsf{EP}_{1},\textsf{EP}_{2}\}$.
Lifting \(X\) to its two-sheeted Riemann surface produces
\(Y=S_r\setminus\{RP_1,RP_2\}\), where $S_{r}$ is a capped Riemann Surface and each ramification point
\(RP_j\) lies above the corresponding branch point \((\textsf{EP})_j\).
The canonical covering map
\(\phi^{\circ}:Y\to X\) satisfies \(\phi^{\circ}(RP_j)=(\textsf{EP})_j\)
\((j=1,2)\) and is a local homeomorphism away from the RPs.
This covering induces a homomorphism on fundamental groups,
\begin{align}
(\phi^{\circ})_{*}:\;
\pi_{1}(Y,y_{0})
&\;\longrightarrow\;
\pi_{1}(X,x_{0}), \\[4pt]
[\tilde{\alpha}]
&\;\longmapsto\;
[\phi^{\circ}\!\circ\!\tilde{\alpha}],
\end{align}
with \(\phi^{\circ}(y_{0})=x_{0}\).
Its image,
\((\phi^{\circ})_{*}\bigl(\pi_{1}(Y,y_{0})\bigr)
     \le \pi_{1}(X,x_{0})\),
is a subgroup of the loop group on the base space.
Hence lifting a loop
\([\alpha]\in\pi_{1}(X,x_{0})\) to
\([\tilde{\alpha}]\in\pi_{1}(Y,y_{0})\)
fully captures the non-Abelian winding behaviour around the EP pair.

We first consider the homotopy class of a loop $[\alpha] \in\pi_{1}(X,x_{0})$.
Because the base space $X$ is a twice-punctured plane, it is homotopy equivalent to the figure eight, $X\simeq S^{1}\vee S^{1}$.
Hence its fundamental group is isomorphic to the free group with two generators,
$\pi_{1}(S^{1}\vee S^{1})\cong F_{2}\cong\langle a,b\rangle$,
where $a$ and $b$ wind around $\textsf{EP}_1$ and $\textsf{EP}_2$, respectively.
Any loop $[\tilde{\alpha}] \in \pi_{1}(Y,y_{0})$ can therefore be expressed in the alphabet $\langle a,b\rangle$.

Each panel in the upper part of Fig.~\ref{Figure-2}(a)-(h) displays a non-trivial loop on the Riemann surface, $[\tilde{\alpha}]\in\pi_{1}(Y,y_{0})$, together with its projection on the base space, $[\alpha]=(\phi^{\circ})_{*}([\tilde{\alpha}])\in\pi_{1}(X,x_{0})$.
Panel (a) shows a loop encircling $RP_1$; its projection is the two‑fold loop $[\alpha]=a^{2}$ around $\textsf{EP}_1$ on the real sheet.
Panel (e) shows the corresponding loop on the imaginary sheet.
Panel (b) illustrates a loop encircling $RP_2$, whose projection is $[\alpha]=b^{2}$ around $\textsf{EP}_2$ (real sheet); panel (f) gives its imaginary sheet counterpart.
Panel (c) shows a loop encircling $RP_1$ and then $RP_2$ individually; the projected loop is $[\alpha]=ab$ around $\textsf{EP}_1$ and $\textsf{EP}_2$ (real sheet).
Panel (g) presents the imaginary‑sheet counterpart; note that the visible ordering on the plane is $ba$, not $ab$.
Panel (d) depicts a loop encircling $RP_1$ and $RP_2$ simultaneously; its projection is $[\alpha]=c$ around both \textsf{EP}s (real sheet), with panel (h) showing the same for the imaginary sheet.

Moreover, the loop $ba$ is homotopic to the loop the $c$ under the homotopy $H:[0,1]\times[0,1]\longrightarrow X$ such that $H|_{[0,1]\times 0}=\alpha_{0}$ is a representative of $ba$ and $H|_{[0,1]\times 1}=\alpha_{1}$ is a representative of $c$. (See the details in supplementary). Therefore, the loop $[\tilde{\alpha}]$ on Riemann surfaces in Fig. Fig.~\ref{Figure-2}(c) are homotopic to the the loop $[\tilde{\alpha}]$ on Riemann surfaces in Fig.~\ref{Figure-2}(d). More precisely,
 if $H$ is a homopoty with path lifting $\tilde{\alpha_{0}}$ of $\alpha_{0}$=$H|_{[0,1]\times 0}$, there exists a homotopy $\tilde{H}:[0,1]\times [0,1]\longrightarrow Y$ lifting $H$ with $\tilde{\alpha_{0}}=\tilde{H}|_{[0,1]\times 0}$ and $\tilde{\alpha_{1}}=\tilde{H}|_{[0,1]\times 1}$ as shown below:
\begin{equation}\label{eq:homotopy-lift}
\xymatrix{
  I \ar[r]^{\tilde{\alpha}_0} \ar[d]_{I\times\{0\}} &  Y \ar[d]^{\phi^{\circ}} \\
  I \times I \ar[r]_H \ar@{..>}[ru]^{\tilde{H}} & X
}
\end{equation}
Thus the lifted loops in Fig.~\ref{Figure-2}(c) and Fig.~\ref{Figure-2}(d) are homotopic in $Y$.
Analogously, the loop $ab$ is homotopic to the loop $d$ shown in Fig.~\ref{Figure-2}(j).
Figures~\ref{Figure-2}(i) and \ref{Figure-2}(j) depict the homotopy lifts for the CW and CCW loops $c$ and $d$, respectively.

The set $\{a^{2},b^{2},ab\}$ forms a free generating set for $\pi_{1}(Y,y_{0})$.
For example, $ba$ can be written as $ba=b^{2}(ab)^{-1}a^{2}$.
Consequently, every loop on $Y$ based at $y_{0}$ can be written uniquely as a reduced word in $\{a^{2},b^{2},ab\}$.

Two full traversals around a single \textsf{EP}, such as $a^{2}$ or $b^{2}$, are required for the eigenvalues $\lambda_{r}$ and $\lambda_{i}$ to return to their original sheets.
By contrast, a single traversal of loop $c$ or $d$ suffices when both branch points are enclosed.
These behaviours are characteristic of the square-root surface $f(z)=\sqrt{(z-z_{1})(z-z_{2})}$, as illustrated in Fig.~\ref{Figure-1}(a-f).

\section{Equivalence between Hopping and gluing of complex planes }
\begin{figure}
  \centering
  \includegraphics[width=8.8cm]{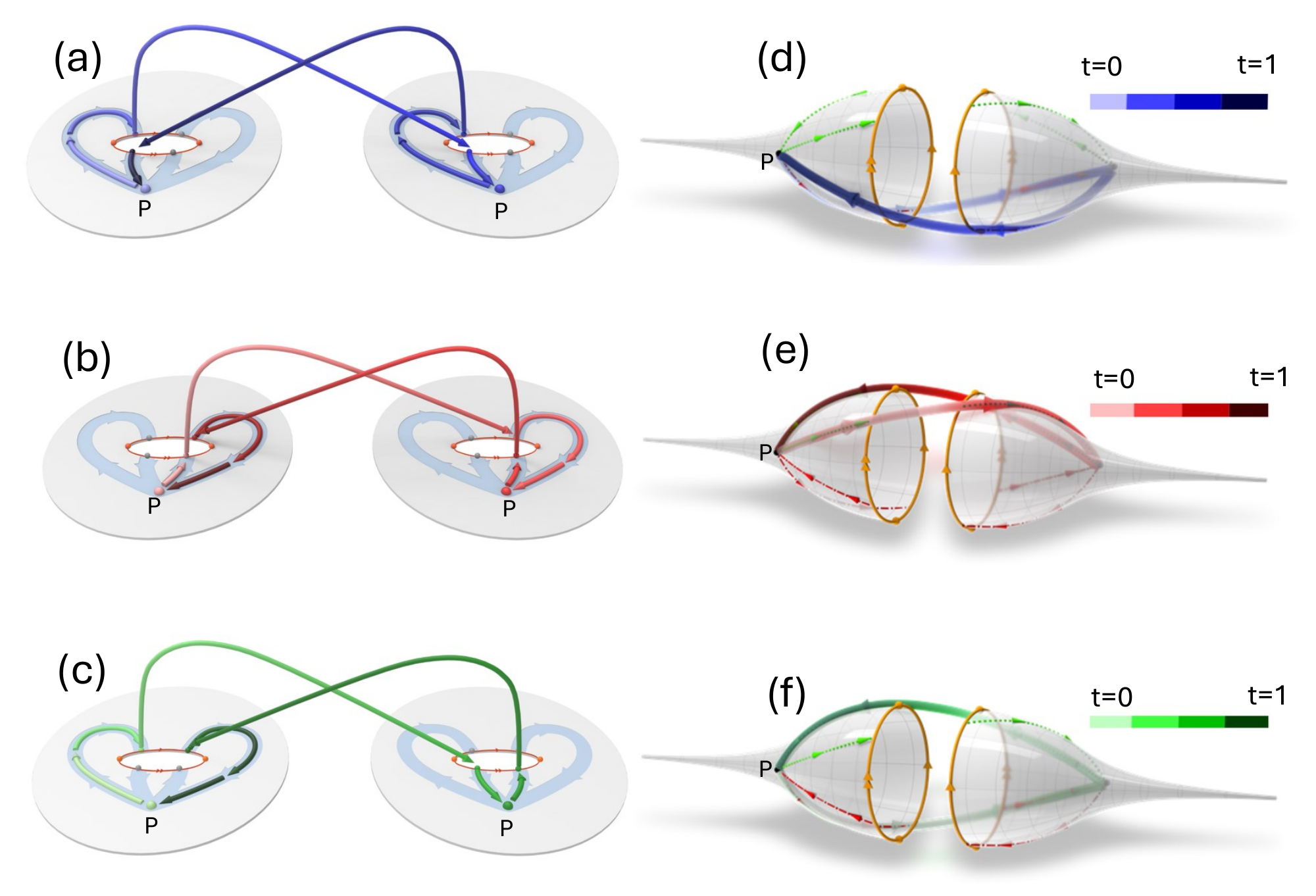}
  \caption{(a-c) Hopping picture: two copies of the punctured plane
  \(\mathbb{C}\setminus\{-i,i\}\) arranged side-by-side.
  Two hops across the branch cut at \(-i\) realise the loop \(a^{2}\) (a);
  two hops across \(i\) realise \(b^{2}\) (b);
  one hop across each cut realises \(ab\) (c).
  (d-f) Gluing construction that corresponds to the hopping moves in (a-c),
  demonstrating their one-to-one correspondence.
  Colored arrows indicate sheet changes along the loop from \(t=0\) to \(t=1\).}
  \label{Figure-3}
\end{figure}

To readily identify path homotopies on our two-sheeted Riemann surfaces and to extend this construction from two sheets with two \textsf{EP}s to an arbitrary
$n$-sheeted surface with $2$m \textsf{EP}s, we introduce an alternative, but homeomorphic, representation of these surfaces that makes their sheet structure and branch cuts immediately transparen. To this end, we observe that when crossing a branch cut from a Riemann sheet, it lands on the other Riemann sheet. From this observation, we can conjecture that hopping between two copies of punctured complex planes can be equivalent to the loop of Riemann surfaces, i.e., figures (a-c) in Fig.~\ref{Figure-3} clearly correspond to the loops $a^{2}$, $b^{2}$, $ab$ in Fig.~~\ref{Figure-2}, respectively.  Then, we consider gluing the two copies of punctured complex planes as follows. First, cut the $\CC$ along the arc $\alpha$ connecting $\textsf{EP}_1(=-i)$ and $\textsf{EP}_2(=i)$  resulting in surfaces that are homeomorphic to  $\CC\setminus \DD$ where $\DD=\{z\in \CC:|z|<1\}$. Then, take two copies of $\CC\setminus \DD$ and place them opposite side with respect to mirror anti-symmetry, finally glue them along the boundary $\partial \DD$ with respect to $-i$ and $i$.  Namely,
\begin{align}
S=\{(\CC\setminus \DD)
\times \ZZ/2\ZZ \}/ \sim
\end{align}
where the attaching relation $\sim$ is defined as $(u+v i,0)\sim (-u+v i,1)$ for any $u+v i\in \partial \DD$.  Then, we can figure out that the loops of $S$ in Fig.~\ref{Figure-3} (d-f) is equivalent to the loops of Fig.~\ref{Figure-2} (The Riemann surface in Fig.~\ref{Figure-2} and surface $S$ are homotopy equivalent). Moreover, we easily find out that they are also homeomorphic to each other. The attaching relation $\sim$ also can be realized with respect to mirror symmetry such that attaching must take place crossing the each branch cut, and we immediately  notice that this topological space is also homeomorphic to the Riemann surface in Fig.~\ref{Figure-2}. As a result we obtained twice punctured Riemann sphere. From the construction, we can see that each puncture of $S$ comes from the infinity of $\CC$.

\section{Topological chirality}
\begin{figure*}
  \centering
  \includegraphics[width=\textwidth]{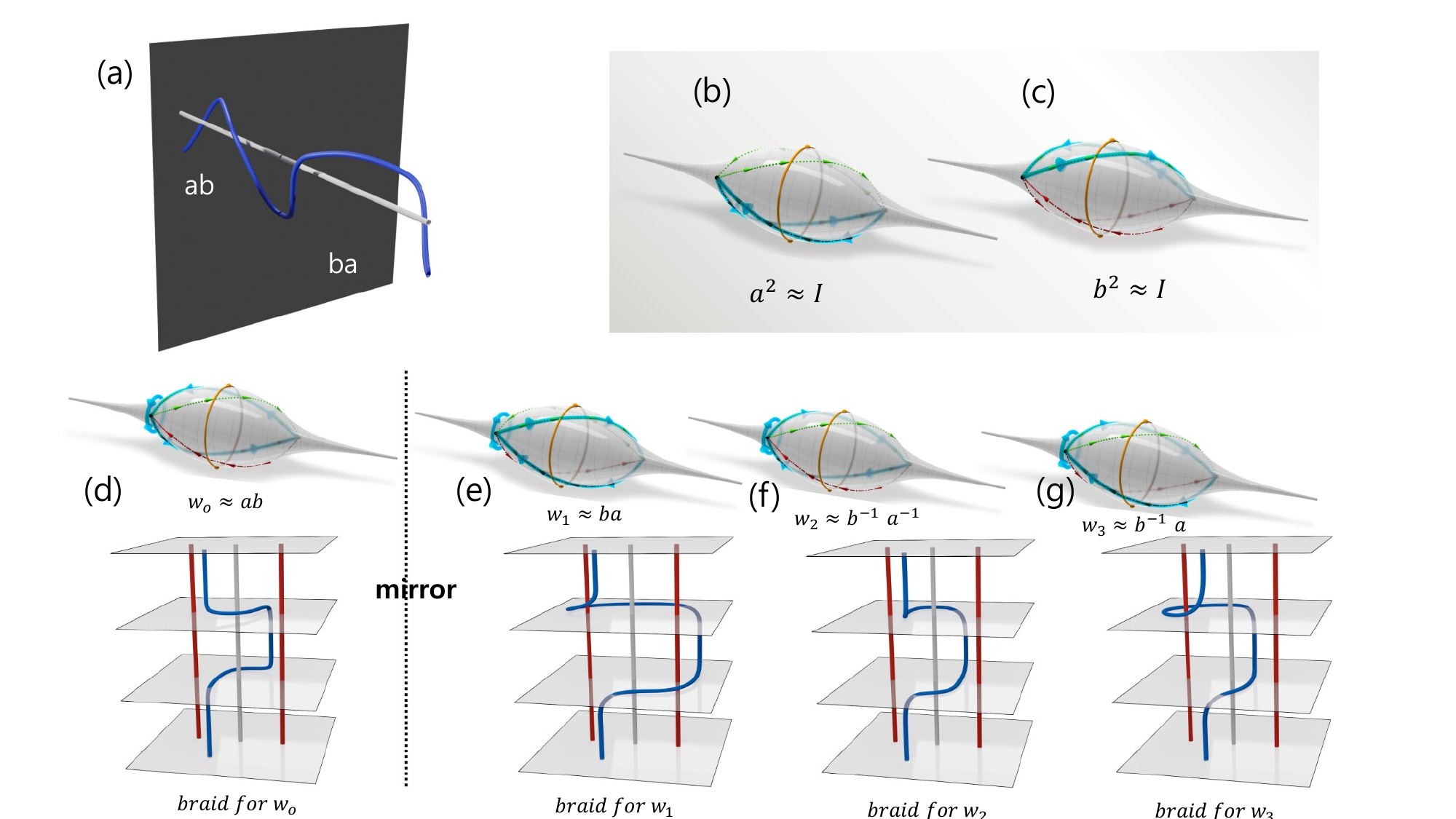}
  \caption{
Topological chirality for a pair of \textsf{EP}s. (a) Braid for the loop \(w_{0}=ab\) concatenated with its mirror-image inverse. When the crossings in the concatenated braid are pulled straight, they cancel out and reduce to the trivial braid (two parallel strands).
Loops (b) and (c) on the Riemann surface are homotopically trivial, since they shrink to the identity once the cone relations \(a^2 = b^2 = e\) are imposed. . Panel (d) shows the clockwise loop $w_{0} = ab$ and its braid, while panels (e), (f), and (g) display three counter-clockwise mirror loops $w_{1} = ba$, $w_{2} = b^{-1}a^{-1}$, $w_{3} = b^{-1}a$ and their braid, respectively. More generally, given any reference loop \(w_{0}\) (or its corresponding braid), one obtains two natural categories: the set of all loops homotopic to \(w_{0}\), and the \emph{topological chirality set} \(\{w_{\alpha}\}\), whose mirror-image elements satisfy $w_{0}w_{\alpha} = e$ in the capped fundamental group.}
  \label{Figure-4}
\end{figure*}

\paragraph*{Fixed points and orbifold singularities.}
\textsf{EP}s in non-Hermitian systems naturally arise as \emph{fixed points} under a group action that builds a covering quotient. Specifically, if a group \(G \cong \mathbb{Z}_2\) acts on a two-sheeted Riemann surface \(Y\) by swapping the sheets (\(z,+\mapsto z,-\)), then any point fixed by this operation---such as \(z=0\) in \(\sqrt{z}\)---emerges as a \emph{cone point} in the quotient space \(X = Y/G\). These fixed points do not admit smooth neighborhoods in \(X\); instead, they produce \emph{orbifold singularities} modeled on \(\mathbb{R}^2/\mathbb{Z}_2\). In physical terms, each cone point corresponds to an \textsf{EP}: encircling it once moves the system’s state to a different state, and only after two loops does one return to the original branch, encapsulated by \(a^{2}=e\) (with \(a\neq e\)). Thus, the essential two-sheet topology of \(Y\) is effectively embedded within the experimentally tunable parameter space.

\paragraph*{Orbifold group presentation.}
When two \textsf{EP}s arise, placing order-2 cone points at those locations yields the orbifold sphere
\(
  S^{2}(2,2,\infty)\).
Let \(a\) and \(b\) denote loops winding once around \(\textsf{EP}_{1}\) and \(\textsf{EP}_{2}\), respectively. Imposing the cone-point constraints \(a^{2} = b^{2} = e\) means factoring out the smallest normal subgroup \(\langle\!\langle a^{2},\,b^{2}\rangle\!\rangle\)  containing \(a^{2}\) and \(b^{2}\) from the free group \(\langle a,b\rangle\). Hence
\begin{align}
  \pi_{1}^{\mathrm{orb}}\bigl(S^{2}(2,2,\infty)\bigr)
  \;\cong\;
  \langle a,b\rangle
  \;\Big/\; \langle\!\langle a^{2},\,b^{2}\rangle\!\rangle
\end{align}
where \(\langle\!\langle a^{2}, b^{2}\rangle\!\rangle\) is the smallest normal subgroup of \(\langle a,b\rangle\). This presentation, featuring two generators and order-2 relations, defines the infinite dihedral group $D_\infty$, capturing the algebraic structure for \textsf{EP}-induced branch switching and forming the basis of our topological chirality analysis.

\paragraph*{Cone relations as topological capping.}
This abstract group structure admits a clear geometric interpretation. Imposing \(a^2 = e\) and \(b^2 = e\) is effectively equivalent to “capping” the corresponding ramification points on \(Y\), producing the capped Riemann surface \(S_r\). As a result, some loops on \(S_r\) become contractible.
 In Fig.~\ref{Figure-4}(b) and (c), for instance, the loops become homotopically trivial, namely, it can shrink to a point. By contrast, the loop in panel (d) cannot contract, as it winds nontrivially around the starting point $P$; we label this a clockwise (CW) loop. Likewise, panels (e), (f), and (g) show three mirror loops in the counter clockwise (CCW) sense.

\paragraph*{Chirality and non-commutativity.}
Crucially, these CW and CCW loops are \emph{not} homotopic to each other: no continuous deformation can turn one into the other. This obstruction reflects the non-Abelian nature of the orbifold group: in particular, $ab \ne ba$. We thus define  \emph{topological chirality} as the inequivalence of CW and CCW encirclements triggered by the non-commuting loop classes. Unlike in knot theory, where (for instance) the figure-eight knot is deformable into its mirror, the \textsf{EP}-driven chirality observed here is protected by the orbifold fundamental group.

\begin{table*}[t]
\centering
\caption{
Reduced words, EP crossings, linking numbers, and vorticity for $z=1,2$ and general $z=k$ (CW).
Vorticity is defined as $\nu(\Gamma) = \sigma_p / \mathrm{ord(EP)}$, where
$\sigma_p = \sum p_i$ with $p_i = \pm1$ from reduced word powers, and $\mathrm{ord(EP)} = 2$.
Thus, for word length $2k$, $\nu(\Gamma) \in \{-k, -k+1, \dots, k\}$, following a symmetric binomial distribution.
}
\small
\begin{tabular}{c|c|p{5.5cm}|c|c}
\hline
\multicolumn{5}{c}{\bfseries $z=1$ ($k=1$, word length $2$)}\\
\hline
RP(EP) crossings $r$ & \# reduced words & Examples & Linking number $Lk=1+r$ & $\nu(\Gamma) = \sigma_p / \mathrm{ord(EP)}$ \\
\hline
0 & 1 & $ab$ & 1 & 1 \\
1 & 2 & $a^{-1}b,\;ab^{-1}$ & 2 & 0 \\
2 & 1 & $a^{-1}b^{-1}$ & 3 & -1 \\
\hline
\multicolumn{5}{c}{Total reduced words: $2^2 = 4$} \\
\hline\hline

\multicolumn{5}{c}{\bfseries $z=2$ ($k=2$, word length $4$)}\\
\hline
RP(EP) crossings $r$ & \# reduced words & Examples & Linking number $Lk=2+r$ & $\nu(\Gamma) = \sigma_p / \mathrm{ord(EP)}$ \\
\hline
0 & 1 & $abab$ & 2 & 2 \\
1 & 4 & $a^{-1}bab$ & 3 & 1 \\
2 & 6 & $a^{-1}b^{-1}ab$ & 4 & 0 \\
3 & 4 & $a^{-1}b^{-1}a^{-1}b$ & 5 & -1 \\
4 & 1 & $a^{-1}b^{-1}a^{-1}b^{-1}$ & 6 & -2 \\
\hline
\multicolumn{5}{c}{Total reduced words: $2^4 = 16$} \\
\hline\hline

\multicolumn{5}{c}{\bfseries General $z = k$ ($k \ge 1$, word length $2k$)} \\
\hline
RP(EP) crossings $r$ & \# reduced words & Comment & Linking number $Lk = k + r$ & $\nu(\Gamma) = \sigma_p / \mathrm{ord(EP)}$ \\
\hline
$r=0,1,\dots,2k$ & $\binom{2k}{r}$ & Invert $r$ out of $2k$ segments to set EP crossings to $r$ & $k + r$ & $\nu(\Gamma) \in \{-k, -k+1, \dots, k\}$ \\
\hline
\multicolumn{5}{c}{Total reduced words: $2^{2k}$} \\
\hline
\end{tabular}
\end{table*}

\paragraph*{Capped Riemann surface and loop reduction.}
To confirm these results, we can compute the fundamental group of the constrained (“capped”) Riemann surface. Once $a^2 = e$ and $b^2 = e$ hold, the only surviving generator corresponds to $ab$, giving
\begin{equation}
  \pi_1(S_r)
  \;\cong\;\frac{\langle a^2,b^2,ab\rangle}{\langle\!\langle a^2,b^2\rangle\!\rangle}
  \;=\;\langle ab\rangle
  \;\cong\;\mathbb{Z}.
\end{equation}
More precisely, this follows from the First Isomorphism Theorem:
$G/\ker\varphi \;\cong\; \langle ab\rangle \;\cong\; H$,
where $G=\langle a^2,b^2,ab\rangle$, $H=\mathbb{Z}$, and
$\ker\varphi=\langle\!\langle a^2,b^2\rangle\!\rangle$.

Consequently, $\pi_1(S_r)$ is isomorphic to the circle's fundamental group $\mathbb{Z}$.  In physical terms, all non-trivial loops on this capped surface reduce to CW or CCW windings around both \textsf{EP}s.  For a single CW or CCW winding we obtain four reduced words (loops) in each case: $\varphi^{-1}(1)=\{ab,a^{-1}b,ab^{-1},a^{-1}b^{-1}\}$ for CW and $\varphi^{-1}(-1)=\{ba,b^{-1}a,ba^{-1},b^{-1}a^{-1}\}$ for CCW.

Defining the parity operator $P$, which reverses the orientation of any loop, we have
\[
  P:\pi_1(S_r)\longrightarrow\pi_1(S_r),\qquad P([\gamma])=[\gamma]^{-1}.
\]
Under the identification $\pi_1(S_r)\cong\mathbb{Z}$ this becomes $P(n)=-n$, so $P$ exchanges the CW set $\{ab,a^{-1}b,ab^{-1},a^{-1}b^{-1}\}$ with the CCW set $\{ba,b^{-1}a,ba^{-1},b^{-1}a^{-1}\}$.

We now adopt an \emph{extended homotopy} convention that allows a loop to slide across an EP (or its real projection~RP).  With this broader equivalence, Fig.~\ref{Figure-3}(d–g) illustrates how each CCW word contracts: the loop $ba$ in Fig.~\ref{Figure-3}(e) must cross both RP1 and RP2 to shrink to the CCW loop at~P; the loop $b^{-1}a$ in Fig.~\ref{Figure-3}(f) crosses RP1 once; the loop $b^{-1}a^{-1}$ in Fig.~\ref{Figure-3}(g) crosses no RPs; and, although not shown, the loop $ba^{-1}$ would likewise contract after crossing RP1 once.

As an immediate illustration of these combinatorial patterns, for $z=2$ ($k=2$, word length 4) one finds EP crossings $r=0,\dots,4$ with $\binom{4}{r}$ reduced words and linking number $\mathrm{Lk}=2+r$, and in general for $z=k$ ($k\ge1$, word length $2k$) there are $\binom{2k}{r}$ words with $r$ crossings and $\mathrm{Lk}=k+r$.

This combinatorial freedom is more than a numerical curiosity: it represents a qualitative leap in how EP topology can be engineered.
Traditional EP–encircling studies focus on \emph{how many} times a loop winds around the singularities; here, by selectively flipping the orientation of individual segments, we \textbf{design the exact number of EP crossings itself}.   Such segment-level programmability unlocks a whole spectrum of topological states whose cardinality grows exponentially ($2^{2k}$) with the base winding $k$,


Our argument can be further clarified by investigating the relationship between the fundamental group and the braid group, since the inverse of any braid is given by its mirror image.
To do this, we need to introduce some generalization of braid groups.
A usual braid is presented as a tangled $n$-strands.
We may interpret a braid as a trajectory of $n$-distinct points on the plane or a disk, allowing some perturbation of the trajectory.
In general, this interpretation allows us to extend the concept of braids to general surfaces or topological spaces.

More precisely, given a (topological) space $T$, for $n>0$, we define the \emph{configuration space} $X_n(T)$ of $n$-distinct points as follows:
\[
X_n(T)=\{(t_1,t_2,\cdots, t_n)\in T\times \cdots \times T: t_i\neq t_j\text{ for all }i\neq j\}
\]
Then, the \emph{$n^{th}$-braid group} $B_n(T)$ in  $T$ is defined as the fundamental group of $X_n(T)/S_n$ with some base-point $t\in X_n(T)/S_n$.
Here, $S_n$ is the group of coordinate permutations.
The interesting part is that the first braid group $B_1(T)$ is exactly the fundamental group of $T$ since $X_1(T)=T$.
In topology, the most significant case arises when $T$ is a surface.
In this case, the braid groups are called \emph{surface braid groups}. See \cite{B74} for a detailed exposition of the theory.

To see that the topological chirality set comes from the mirror images of $w_0$, we think of the fundamental group of the Riemann surface $S_r$ as the braid group of $S_r$ with only one strand. For instance, see Fig~\ref{Figure-4}-(d).
Here, each plane presents some moment of the surface.
The bottom plane is the surface at time $0$ and the top plane is the surface at time $1$.
Two red strands are the trajectory of the $RP_1$ and $RP_2$ which are not actual strands. The white strand shows the trajectory of the left cusp of the surface in Fig~\ref{Figure-4}-(c).
This plays a role as a strand.
The blue strand is the trajectory of the point $P$ in Fig~\ref{Figure-4}-(c). Hence, the blue strand, together with the white strand, forms a braid with corresponding to $ab$.
The red strands are the trajectory of the RPs. Likewise, Fig~\ref{Figure-4}-(e), is the braid corresponding to $ba$. As displayed in Fig~\ref{Figure-4}-(a), we can observe that the mirror image of $ab$ coincides with the braid of $ba$. Therefore, the product of those gives rise to the trivial braid.
From Fig~\ref{Figure-4}-(f) and  Fig~\ref{Figure-4}-(g), it follows that the braids of $b^{-1}a^{-1}$ and $b^{-1}a$ provide same braid which equal to that of $ba$.

\paragraph*{Generalization to multiple \textsf{EP}s and surfaces.}
More generally, for any chosen reference loop (word) \(w_{0}\) on a two-sheeted Riemann surface (or, equivalently, its corresponding braid), two distinct categories are naturally defined:
\begin{itemize}
  \item Loops homotopic to \(w_{0}\).
  \item The \emph{topological chirality set} \(\{w_{\alpha}\}_{\alpha}\), whose elements are mirror-image loops satisfying \(w_{0}w_{\alpha}=e\) in the capped fundamental group.
\end{itemize}
Crucially, this construction is not confined to the simple two-\textsf{EP} case above. The same orbifold fundamental-group and braid-group arguments extend to Riemann surfaces of arbitrary sheet number (for example, \(n\) sheets) with any even number \(2m\) of \textsf{EP}s.


\section{Conclusions}
We introduce a new, strictly topological form of chirality, called \emph{topological chirality} that emerges whenever a system supports an exceptional-point (\textsf{EP}) pair. Conventional non-Hermitian manifestations of chirality, such as mode imbalance, energy-flow vortices, and encircling-induced holonomy, are not topological invariants and may change under smooth variations of system parameters. In contrast, the topological chirality defined here is intrinsic and robust: it is protected by the non-commutative structure of the orbifold fundamental group and its associated braid group, and it remains invariant under continuous deformations of the Hamiltonian or the encirclement path.

Because of this protection, every loop belongs to one of two inequivalent classes (a loop and its mirror image), and the distinction yields physically measurable consequences: a loop and its mirror image exhibit opposite spectral vorticity, conjugate complex Berry phases, and distinct permutations in the non-Abelian holonomy matrix. We demonstrate these predictions using two physically distinct platforms, an optical microcavity and a non-Hermitian Dirac lattice, and the same two-sheeted Riemann-surface topology reappears in classical spin systems, photonic crystals, and hybrid light--matter structures, underscoring the generality of our framework.

Moreover, our ``gluing-of-planes'' construction (Sec.~III), grounded in rigorous orbifold and braid theory, naturally extends the concept to \(n\)-sheeted Riemann surfaces with \(2m\) \textsf{EP}s, ensuring that topological chirality remains meaningful in more elaborate configurations. These results refine the theoretical foundations of non-Hermitian physics and open new directions for experimental and applied research: because the chiral response is topologically protected, it furnishes a robust and versatile design principle for next-generation sensors, non-reciprocal waveguides, and other asymmetric transport devices.


\section{Acknowledgments}
This work was supported by the National Research Foundation of Korea (NRF) through a grant funded by the Ministry of Science and ICT (Grants Nos. RS-2023-00211817, RS2022-NR072395, RS-2022-NR068791, and RS-2025-00515537), the Institute for Information \& Communications Technology Promotion (IITP) grant funded by the Korean government (MSIP) (Grants Nos. RS-2019-II190003 and RS-2025-02304540), the National Research Council of Science \& Technology (NST) (Grant No. GTL25011-000), and the Korea Institute of Science and Technology Information (KISTI).

\bigskip
\section*{Disclosures}
The authors declare no conflicts of interest.

\section*{Author Contributions}
K.-W.P. proposed the study and performed theoretical analysis. K.K. performed the theoretical and mathematical analysis. K.J. supervised the overall investigation. K.-W.P., K.K., and K.J. wrote the manuscript. K.-W.P. and K.K. contributed equally to this work.

\section{Supplementary}
\makeatletter
\setcounter{figure}{0}            
\renewcommand{\thefigure}{S\arabic{figure}}  
\makeatother

\subsection{Riemann sphere for understanding of topological structure of double \textsf{EP}s}
The Riemann sphere is a mathematical model that describes an extended complex plane. The extended complex plane comprises a complex plane ($\mathbb{C}$) and a point at infinity, which is represented by $\mathcal{C}_{\infty}=\mathbb{C}\cup{\{\infty\}}$. In this section, we exploit this extended complex plane ($\mathcal{C}_{\infty}$) and Riemann sphere to understand the topological structure of double \textsf{EP}s. For this analogy, we consider an isomorphism between the parameter plane $\mathbb{P}=\{(n,\chi)|n, \chi\in\mathbb{R}\}$ and a complex plane $\mathbb{C}$: $\mathbb{P}$ $\cong$ $\mathbb{C}$ defined by ($n,\chi$)$\mapsto$ $n+\iota\chi$. Thus, we can naturally adopt an extended parameter space to utilize a Riemann sphere: $\mathbb{P}_{\infty}$=$\mathbb{P}\cup{\{\infty\}}\cong\mathcal{C}_{\infty}$.

As described in Sec. III, we cannot analytically obtain the eigenvalues ($\lambda_{i}$). However, we can numerically calculate $\lambda_{i}$ as a function of only two parameters ($n,\chi$). Accordingly, we can adopt a functional map ($\widehat{\lambda_{i}}$) from $\mathbb{P}_{\infty}$ to $\mathcal{C}_{\infty}$:
\[
\widehat{\lambda_{i}}:\mathbb{P}_{\infty}\to \mathcal{C}_{\infty}
\label{eq8}
\]

However, we considered only two specific eigenstates and neglected the other states, but this assumption is only possible within a restricted parameter plane. This is because the two specific eigenstates can interact with other eigenstates in other ranges in the parameter plane. Moreover, infinity cannot be dealt with in a practical physical system. Therefore, we only performed numerical simulations of the restricted range in the parameter plane, as shown in Fig. ~\ref{Figure-1}(a-b), where $p=[n_{{\rm in}; 1}, n_{{\rm in}; 2}]\times[\chi_{1}, \chi_{2} ]\subset \mathbb{P}_{\infty}$ with $n_{{\rm in};1}=2.6, n_{{\rm in}; 2}=3.0$ and $\chi_{1}=0.50, \chi_{2}=0.63$, respectively. Thereby, we can define a restricted map $\widehat{\lambda_{i}}|p$: $p \to \mathbb{C}$ with the condition $\lambda_{1}(n,\chi)=\lambda_{2}(n,\chi)$) if and only if $(n,\chi)\in\{\textsf{EP}1, \textsf{EP}2\}$.

In order to address the structures of branch cuts that can go to infinity, we should extend the restricted map $\widehat{\lambda_{i}}|p$ to the $\widehat{\lambda_{i}}$ under the certain conditions: (a) $\widehat{\lambda_{i}}((n,\chi)$=$\widehat{\lambda_{i}}|{p}(n,\chi)$ for all $(n,\chi)\in p$; (b) $\widehat{\lambda_{1}}(n,\chi)=\widehat{\lambda_{2}}(n,\chi)$) if and only if $(n,\chi)\in\{\textsf{EP}1, \textsf{EP}2\}$; (c) $\widehat{\lambda_{i}}(n=\infty,\chi=\infty)<\infty$. Under these conditions, our $2\times2$ non-Hermitian Hamiltonian is well defined for the double \textsf{EP}1, \textsf{EP}2s in the extended parameter plane $\mathbb{P}_{\infty}$, and can yield branch cuts at infinity.

We then exploit the Riemann sphere to understand the topological structure of double \textsf{EP}1, \textsf{EP}2s by stereographic projection. Stereographic projection is a smooth and bijection map that projects a sphere onto a plane. In this study, the Riemann sphere is an embedded manifold in a parameter $\mathbb{R}^{3}$ space that comprises of ($\tilde{n}, \tilde{\chi}, \tilde{\xi}$). Here, $\tilde{\xi}$ is a dummy axis for building up $\mathbb{R}^{3}$ space. Thus, we can consider a unit sphere $\mathbb{S}^{2}$ as a Riemann sphere in the parameter space $\mathbb{R}^{3}$: $\mathbb{S}^{2}=\{(\tilde{n}, \tilde{\chi}, \tilde{\xi})\in \mathbb{R}^{3}$: $\tilde{n}^{2}+\tilde{\chi}^{2}+\tilde{\xi}^{2}=1\}$ and a stereographic projection map $\Theta : \mathbb{S}^{2}\to \mathbb{P}_{\infty}$.

The stereographic projection of $\mathbb{S}^{2}$ is performed from the north pole $\hat{N}=(0, 0, 1)$ onto the plane $\tilde{\xi}=0$ and is explicitly defined as follows:
\begin{align}
\Theta(\tilde{n}, \tilde{\chi}, \tilde{\xi})
  &= \left( \frac{\tilde{n}}{1 - \tilde{\xi}}, \frac{\tilde{\chi}}{1 - \tilde{\xi}} \right) \tag{S1} \label{eq:S1} \\
\Theta^{-1}(n, \chi)
  &= \left( \frac{2n}{\zeta}, \frac{2\chi}{\zeta}, \frac{n^2 + \chi^2 - 1}{\zeta} \right) \tag{S2} \label{eq:S2}
\end{align}

 It is found that the double \textsf{EP}1, \textsf{EP}2 were located at $(n\simeq 2.6257, \chi=0.6001)$ and $(n\simeq 2.9036, \chi=0.5372)$ on $\mathbb{P}_{\infty}$, respectively. Therefore, the double \textsf{EP}s ($\widetilde{\textsf{EP}1}$ and $\widetilde{\textsf{EP}2}$) on $\mathbb{S}^{2}$ corresponding to {\textsf{EP}1} and {\textsf{EP}2} on $\mathbb{P}_{\infty}$ are located at \(\bigl(\tilde n\simeq0.636,\;\tilde\chi\simeq0.145,\;\tilde\xi\simeq0.758\bigr)\)
and \(\bigl(\tilde n\simeq0.598,\;\tilde\chi\simeq0.111,\;\tilde\xi\simeq0.795\bigr)\), respectively. Accordingly, we plotted these four points in Fig.~\ref{Figure-1}(c).

A branch cut is generally a curve that joins two branch points.
In our case, the red solid curve on $\mathbb{S}^{2}$ is the branch cut for the Riemann surfaces of $\lambda_{r}$, which directly connects the double EPs, and is represented by a longer curve on $\mathbb{P}_{\infty}$ by $\Theta(\tilde{n}, \tilde{\chi}, \tilde{\xi})$. Furthermore, the other branch cut for the Riemann surfaces of $\lambda_{i}$ denoted by blue solid curve are connected indirectly detour the north pole ($\hat{N}$). Accordingly, its stereographic projection onto $\mathbb{P}_{\infty}$ reveals features that are quite different from the stereographic projection of the Riemann surfaces of $\lambda_{r}$. When approaching the north pole $\hat{N}$, $\tilde{\xi}$ goes to one, making both $n, \chi$ also go to infinity by Eq.~(\ref{eq:S2}). Note that these two branch cuts on $\mathbb{P}_{\infty}$ meet at infinity, as shown in Fig.~\ref{Figure-1} (c). Consequently, these two different stereographic projections onto $\mathbb{P}_{\infty}$ explain the two different topological structures of the Riemann sheet in Fig.~\ref{Figure-1}(a) and (b).

\subsection{Homotopic theory for covering spaces}

To understand how loops on a multi–sheeted Riemann surface project down to loops in the base parameter plane, we use covering‐space theory and homotopy. Physically, this describes how paths encircling exceptional points (\textsf{EP}1, \textsf{EP}2) lift to continuous trajectories on the branched energy sheets. Mathematically, the covering map
\[
  \phi\colon Y \to X = \C \setminus \{\pm1\}
\]
relates the fundamental group of the two‐sheeted surface \(Y\) (with branch cuts) to that of the punctured plane \(X\). This correspondence underlies our definition of \emph{topological chirality}.

\subsubsection{Path homotopy}

Two loops \(\alpha,\beta\colon [0,1]\to X\) are homotopic if one can be deformed into the other continuously within \(X\). Concretely, a loop encircling a single EP lifts to a two‐turn path on \(Y\), whereas a loop enclosing both EPs at once lifts to a one‐turn path on each sheet. For simplicity, we fix \(\{\mathrm{EP}_1,\mathrm{EP}_2\}=\{1,-1\}\).

The composite loop \([a][b]\), which winds clockwise around \(\mathrm{EP}_1\) then \(\mathrm{EP}_2\), is given by
\[
  \alpha(t) =
  \begin{cases}
    -1 + e^{-4\pi i t}, & 0 \le t \le \tfrac12,\\[6pt]
     1 - e^{-4\pi i(t-\tfrac12)}, & \tfrac12 \le t \le 1.
  \end{cases}
\]
The single‐circuit loop \([c]\) enclosing both EPs simultaneously is
\[
  \beta(t) =
  \begin{cases}
    -2 + 2e^{-4\pi i t}, & 0 \le t \le \tfrac14,\\[6pt]
    -4\,e^{-2\pi i(t-\tfrac14)}, & \tfrac14 \le t \le \tfrac34,\\[6pt]
     2 + 2e^{-4\pi i(t-\tfrac34)}, & \tfrac34 \le t \le 1.
  \end{cases}
\]
We define a piecewise homotopy \(H\colon [0,1]\times[0,1]\to X\) by
\begin{widetext}
\[
  H(t,s)=
  \begin{cases}
    (1-s)\bigl(-1+e^{-4\pi i t}\bigr) + s\bigl(-2+2e^{-4\pi i t}\bigr),
      & 0 \le t \le \tfrac14,\\[6pt]
    (1-s)\bigl(-1+e^{-4\pi i t}\bigr) + s\bigl(-4e^{-2\pi i(t-\tfrac14)}\bigr),
      & \tfrac14 \le t \le \tfrac12,\\[6pt]
    (1-s)\bigl(1-e^{-4\pi i(t-\tfrac12)}\bigr) + s\bigl(-4e^{-2\pi i(t-\tfrac14)}\bigr),
      & \tfrac12 \le t \le \tfrac34,\\[6pt]
    (1-s)\bigl(1-e^{-4\pi i(t-\tfrac12)}\bigr) + s\bigl(2+2e^{-4\pi i(t-\tfrac34)}\bigr),
      & \tfrac34 \le t \le 1.
  \end{cases}
\]
\end{widetext}

One checks \(H(t,0)=\alpha(t)\) and \(H(t,1)=\beta(t)\), proving \(\alpha\) and \(\beta\) are homotopic in the punctured plane as shown in Fig.~S1.

\begin{figure}[ht]
  \centering
  \includegraphics[width=5.5cm]{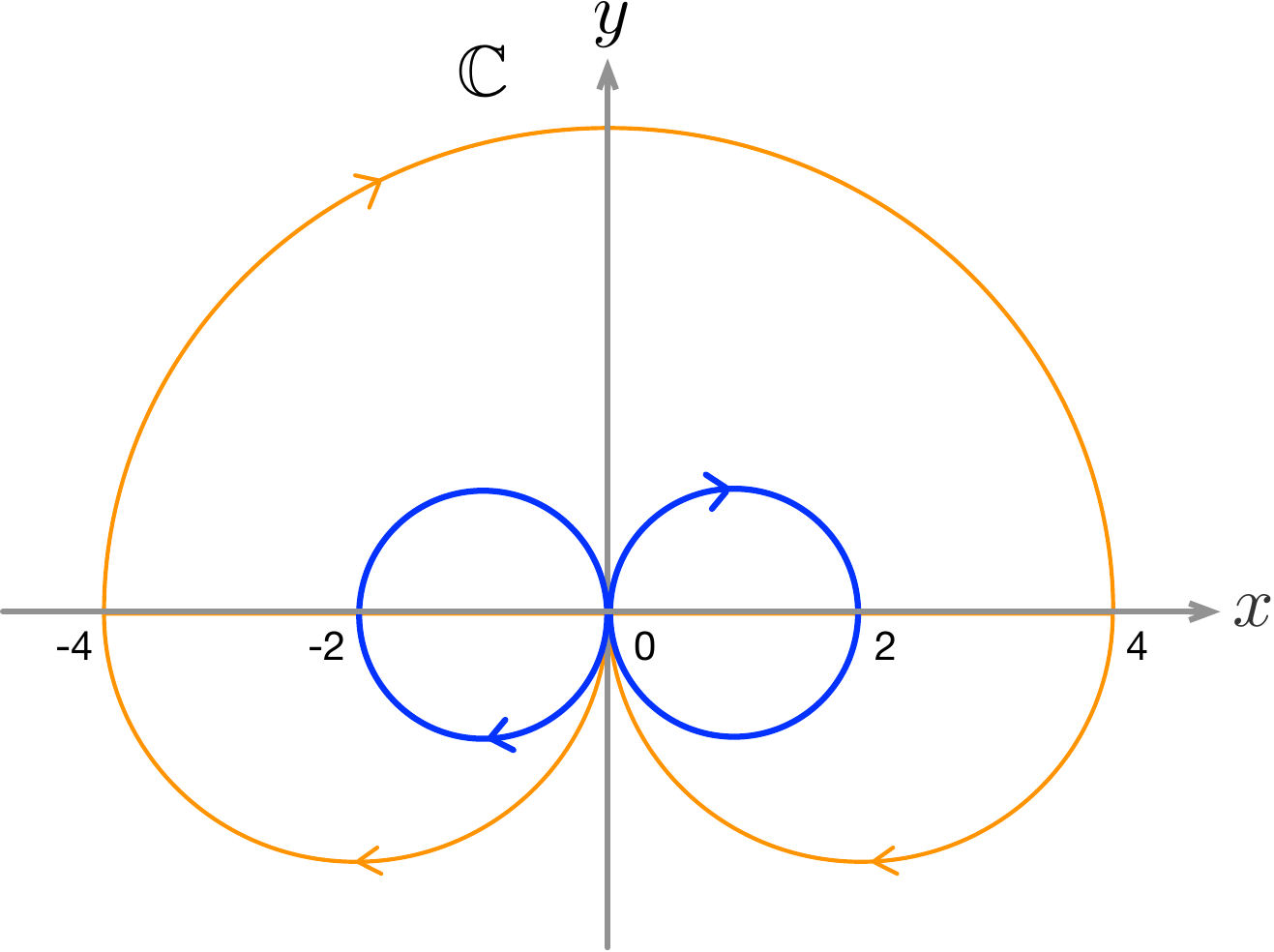}
  \caption{%
    Geometry of the two prototype loops in the complex parameter plane.
    The blue curve \(\alpha(t)\) represents the composite loop \([a][b]\),
    winding clockwise around \(\mathrm{EP}_2=-1\) (inner arc) and then
    \(\mathrm{EP}_1=1\) (outer arc). The orange curve \(\beta(t)\) is a single
    clockwise circuit enclosing both EPs at once. Despite their different
    shapes, these loops are homotopic in \(\C\setminus\{\pm1\}\).}
  \label{fig:S1}
\end{figure}

\subsection*{Riemann Surfaces, Covering Maps, and Orbifold Structure near EPs}
This supplementary material aims to clarify Fig.\ref{conceptDiag}, the conceptual diagram often encountered in the study of multi-valued eigenfunctions near Exceptional Points (EPs).
\begin{figure}[h]
\centering
\scalebox{1.2}{
\begin{tikzcd}[row sep=2em]
\shortstack{\textbf{Riemann Surface } \( Y \) for \( \omega^2=z \)\\with sheets \( +\sqrt{z} \) and \( -\sqrt{z} \)}
\arrow[d, "\text{covering } \phi" right, -{Latex}] \\
\phi : Y \hookrightarrow X = Y/\mathbb{Z}_2
\arrow[d, "\text{quotient }" right, -{Latex}] \\
\shortstack{\textbf{Base Space }$X = \mathbb{C}\setminus\{0\}$\\ with singularity at $z=0$}
\arrow[d, -{Latex}] \\
z=0\text{ is a fixed point }\implies \mathrm{EP}
\arrow[d, -{Latex}] \\
(\textbf{EP} \longleftrightarrow  \text{Cone Point}) \implies \text{Orbifold structure}
\end{tikzcd}
}

\label{conceptDiag}
\end{figure}

Physically, the multi-valued nature of an eigenfunction such as $\sqrt{z}$ indicates that encircling $z=0$ once does not restore the original wavefunction. Mathematically, this arises because $+\sqrt{z}$ and $-\sqrt{z}$ represent two “branches,” and $z=0$ becomes a \emph{branch point} or, in geometric terms, a cone point. Below we break down each component in a manner accessible to physicists with minimal familiarity in topology.

\subsection*{1. Riemann Surfaces and Covering Maps}
\paragraph{Riemann surface $Y$.}
A Riemann surface is, loosely speaking, a “complex 1D manifold” (real dimension 2) that allows \emph{single-valued} definition of otherwise multi-valued functions. For example, $\sqrt{z}$ on $\mathbb{C}\setminus\{0\}$ is two-valued if we stay in the usual complex plane. However, we can build a \textbf{two-sheeted Riemann surface} $Y$ such that each point $(z,\pm)\in Y$ maps to a single $z\in X$ in the base plane.

\paragraph{Covering map $\phi$.}
A covering map $\phi: Y\to X$ is a surjective map where each $x\in X$ typically has the same finite number of pre-images in $Y$. For a 2-sheeted covering, each point of $X$ is “covered” by two points on $Y$, except possibly at branched points.
\[
\phi: (z,\pm) \mapsto z \quad \text{(with branching at $z=0$)}.
\]

In physical examples, $Y$ can represent \emph{two possible eigenvalue/eigenvector branches}. If $z=0$ is a branch point, looping around $z=0$ once in $X$ moves you from one sheet to the other in $Y$ (the essence of a \textbf{nontrivial monodromy}).

\subsection*{2. Group Actions and Quotient Space}
\paragraph{Why $X = Y/\mathbb{Z}_2$?}
Consider the discrete group $G\cong\mathbb{Z}_2 = \{1,g\}$ acting on $Y$ by interchanging the two sheets: $g \cdot (z,+) = (z,-)$. Identifying such pairs means we take a quotient $Y/G$. In simple terms, $G$ toggles the sign of the wavefunction branch. The factor space $X = Y/\mathbb{Z}_2$ can be regarded as “the original complex plane with a branch cut,” i.e., $\mathbb{C}\setminus\{0\}$.

\paragraph{Fixed point $\implies$ singularity.}
When $z=0$ is \emph{fixed} by the action (i.e., $g\cdot(0,+) = (0,-)=(0,+)$), that point yields a \emph{cone} in the quotient. This “cone” is not a smooth manifold feature but a topological singularity. Physically, $z=0$ becomes the branch point (or EP in the non-Hermitian system) around which the wavefunction picks up a sign.

\subsection*{3. EP as a Cone Point and Orbifolds}
\paragraph{Cone point of order 2.}
In 2D, taking $\mathbb{C}/\mathbb{Z}_2$ effectively halves the local angle from $2\pi$ to $\pi$. This is a “cone point of order 2.” Looping once around this point in the quotient is \emph{not} homotopically trivial; you need two loops to return. Symbolically, if $a$ denotes a loop around the point,
\[
a^2 = e \quad (a \neq e)
\]
expresses that “two loops is trivial, one loop is nontrivial.”

\paragraph{Orbifold structure.}
An orbifold is a space that looks locally like $\mathbb{R}^n/G$ for a finite cyclic group $G$. In our case, $G=\mathbb{Z}_2$, so the orbifold has cone points of order 2. The presence of an EP in a non-Hermitian parameter space can be reinterpreted as “the space has a $\mathbb{Z}_2$ cone point at that EP.” If multiple EPs exist, we get multiple cone points, e.g., $S^2(2,2,\infty)$ for a sphere with two order-2 points and one cusp point, which is considered as a point of infinite order.

\subsection*{4. Physical Interpretation: Monodromy and Branch Switching}
\paragraph{Monodromy near EP.}
Physically, an EP merges two eigenmodes. Looping once around it (in “parameter space” of gain/loss, coupling, etc.) typically changes the wavefunction’s sign or branches. This “chiral” or direction-dependent effect is captured by the group relation $a^2=e$, implying that a single loop is nontrivial while two loops restore the original state.

\paragraph{Encircling multiple EPs.}
When two EPs exist, label them $EP_1$ and $EP_2$. Each yields a generator $a$, $b$ in the fundamental group. Then $a^2=b^2=e$ but also $ab$ may fail to commute with $ba$. One obtains a dihedral or non-Abelian group structure, intimately connected to “orbifold fundamental groups.” This explains more complex braiding or “topological chirality” that can arise in multi-EP configurations.

This fundamental group structure, particularly when non-Abelian, forms the basis of non-Hermitian analogs of braid groups. Such braiding behavior underlies recent interest in topologically robust EP-based mode conversions and state control.

\subsection*{6. Summary and Further Remarks}
We have shown how a multi-valued function, exemplified by $\sqrt{z}$, can be recast in terms of a 2-sheeted Riemann surface. The discrete $\mathbb{Z}_2$ action identifies $+\sqrt{z}$ with $-\sqrt{z}$, making $z=0$ a branch point (or an EP in non-Hermitian systems). Locally, this branch point is a cone point of order 2, thus forming an orbifold structure rather than a smooth manifold. Looping once around it implements $a\neq e$, whereas looping twice yields $a^2=e$.

In physical terms, this clarifies why \emph{encircling} an EP once can produce a “mode switch” or sign flip, yet doing so twice returns the initial state. If two or more EPs exist, additional loops and non-commutative relations can arise, opening the door to more intricate braiding phenomena reminiscent of topological quantum systems. Recognizing $z=0$ as a “cone point” thus connects the geometry of branching to the physics of EP encircling in a straightforward but powerful way.


\end{document}